\documentclass[sigconf]{acmart}

\newcommand{\bftab}{\fontseries{b}\selectfont}

\AtBeginDocument{%
  }

\setcopyright{acmcopyright}
\copyrightyear{2025}
\acmYear{2025}
\acmDOI{XXXXXXX.XXXXXXX}

\acmConference[ICBRA '25]{12th International Conference on Bioinformatics Research and Applications}{September 19--21,
  2025}{Prague, CZ}
  
\acmISBN{978-1-4503-XXXX-X/18/06}

\begin{document}

\title{Enhanced Dermatology Image Quality Assessment via Cross-Domain Training}

\author{Ignacio Hernández Montilla}
\email{ignaciohernandez@legit.health}
\orcid{0000-0003-0356-6619}
\affiliation{%
  \institution{Legit.Health}
  \city{Bilbao}
  \country{Spain}
}
\affiliation{
  \institution{University of Deusto}
  \city{Bilbao}
  \country{Spain}
}

\author{Alfonso Medela}
\email{alfonso@legit.health}
\orcid{0000-0001-5859-5439}
\affiliation{%
  \institution{Legit.Health}
  \city{Bilbao}
  \country{Spain}
}

\author{Paola Pasquali}
\email{pasqualipaola@gmail.com}
\orcid{0000-0003-1504-0665}
\affiliation{%
  \institution{Department of Dermatology, Pius Hospital de Valls}
  \city{Valls}
  \country{Spain}}

\author{Andy Aguilar}
\email{andy@legit.health}
\orcid{0000-0003-0618-6179}
\affiliation{%
  \institution{Legit.Health}
  \city{Bilbao}
  \country{Spain}
}

\author{Taig Mac Carthy}
\email{taig@legit.health}
\orcid{0000-0001-5583-5273}
\affiliation{%
  \institution{Legit.Health}
  \city{Bilbao}
  \country{Spain}
}

\author{Gerardo Fernández}
\email{gerardo@legit.health}
\orcid{0009-0006-9721-9974}
\affiliation{%
  \institution{Legit.Health}
  \city{Bilbao}
  \country{Spain}
}

\author{Antonio Martorell}
\email{antonio@legit.health}
\orcid{0000-0003-1378-1590}
\affiliation{%
  \institution{Dermatology Department, Hospital de Manises}
  \city{Valencia}
  \country{Spain}
}
\affiliation{%
  \institution{Legit.Health}
  \city{Bilbao}
  \country{Spain}
}

\author{Enrique Onieva}
\email{enrique.onieva@deusto.es}
\orcid{0000-0001-9581-1823}
\affiliation{%
  \institution{University of Deusto}
  \city{Bilbao}
  \country{Spain}}

\renewcommand{\shortauthors}{Hernández Montilla et al.}

\begin{abstract}
  Teledermatology has become a widely accepted communication method in daily clinical practice, enabling remote care while showing strong agreement with in-person visits. Poor image quality remains an unsolved problem in teledermatology and is a major concern to practitioners, as bad-quality images reduce the usefulness of the remote consultation process. However, research on Image Quality Assessment (IQA) in dermatology is sparse, and does not leverage the latest advances in non-dermatology IQA, such as using larger image databases with ratings from large groups of human observers. In this work, we propose cross-domain training of IQA models, combining dermatology and non-dermatology IQA datasets. For this purpose, we created a novel dermatology IQA database, Legit.Health-DIQA-Artificial, using dermatology images from several sources and having them annotated by a group of human observers. We demonstrate that cross-domain training yields optimal performance across domains and overcomes one of the biggest limitations in dermatology IQA, which is the small scale of data, and leads to models trained on a larger pool of image distortions, resulting in a better management of image quality in the teledermatology process.
\end{abstract}




\received{16 January 2025}
\received[revised]{26 February 2025}
\received[accepted]{26 February 2025}

\maketitle

\section{Introduction}
Over the last years, the use of teledermatology has become widely accepted, especially after the  SARS-CoV-2 (COVID-19) pandemic. From its global validation \cite{vano2011store}, dermatology has benefited from the advancements in remote care, and it has become a broadly accepted and useful communication method between general practitioners and dermatologists \cite{giavina2021dermatologists}\cite{lopez2022teledermatology}.

There are three main types of teledermatology: real-time, or synchronous; store-and-forward, or asynchronous; and hybrid, which combines real-time and store-and-forward. The most utilized method is store-and-forward, in which pictures can be taken either by the healthcare practitioner or the patients themselves. This method has reportedly been considered useful by dermatologists, showing agreement with practitioners that perform conventional in-person visits \cite{lee2018teledermatology}. In store-and-forward dermatology, patients with an inconclusive diagnosis or who require an in-person appointment or referral to a specialized consultation are contacted more quickly, resulting in a reduction of waiting lists.

Dermatology relies on visual inspection to diagnose conditions affecting the skin, hair, nails, and mucous membranes, which makes digital photography an essential tool for many purposes. These include creating a visual record of the disease in patient's history, showing evidence of treatment efficacy, keeping track of changes in a specific lesion, receiving second opinion other practitioners, or educational and scientific purposes \cite{Pasquali2014PhotographyID}. The heavy use of image data in dermatology explains why the emergence of dermatology artificial intelligence (AI) has mostly occurred in the field of computer vision. Thanks to the current skin image databases, AI has been successfully applied to a variety of scenarios such as skin cancer screening, automatic severity assessment \cite{medela2022automatic}\cite{hernandez2023automatic}\cite{mac2024automatic}, and skin pathology recognition \cite{hogarty2020artificial}.

Among other technical limitations, teledermatology is subject to poor image quality, which can limit the effectiveness of consultations as low-quality images may oblige practitioners to ask for new pictures and can eventually lead to referrals to live consultations, unnecessarily increasing the patient's time to receive a diagnosis \cite{paradela2015teledermatology}. Image quality issues also pose a challenge for novel AI-powered image-based diagnostic support tools, which rely on the correct visual content of an image to provide accurate predictions \cite{maier2022image}.

Image Quality Assessment (IQA) is the area of research that develops quality measures to detect distorted and low-quality images. Such measures which can be broadly divided into three approaches. The most classical IQA methods require the original image and the distorted version to make a comparison, which is called full-reference IQA (FR-IQA). Reduced-reference IQA (RR-IQA) measures do not require the original or "pristine" image, but assume that some partial information of it is available. Finally, no-reference or blind IQA measures (NR-IQA) rely exclusively on the distorted image, with no additional information \cite{okarma2019current}. For each approach, the measures can be either hand-crafted, such as BM \cite{de2013new}, SSIM, NIQE, BLIINDS, DIIVINE or BRISQUE \cite{mittal2012no}, or learned by a machine learning model such as HOSA \cite{Xu2016} or DeepBIQ \cite{bianco2018use}. 

To the best of our knowledge, the research on dermatology IQA is sparse and always has been in the no-reference domain due to the predominance of the store-and-forward teledermatology model. There have been recent works that support the use of deep learning models for blind image visual quality assessment, such as TrueImage \cite{vodrahalli2022development}, ImageQX \cite{jalaboi2023explainable}, and DIQA \cite{montilla2023dermatology}. In all cases, the development of domain-specific image datasets and the annotation of IQA labels was required, which is time-consuming.

This raises some questions: \textit{given the plethora of natural IQA datasets available, is it possible to train an IQA model on natural, everyday image quality datasets and use it in the dermatology domain? To what extent is it required to develop domain-specific datasets for dermatology image quality assessment? Can dermatology IQA models benefit from adding image quality databases outside the dermatology domain? Can a single model perform well across all domains?}

In our preliminary work \cite{montilla2023dermatology}, we demonstrated the potential of deep learning models as a tool to evaluate dermatology images in a store-and-forward teledermatology scenario, using a combination of natural and dermatology IQA datasets. However, some aspects remained unexplored, such as the performance of models trained separately on each natural IQA dataset with no additional dermatological fine-tuning, and the performance at higher resolutions.

In the present work, we present a study of model performance on dermatology IQA comparing the usage of natural image datasets with no additional dermatological data to the combination of natural and dermatology images for training. We began by training several deep learning models on natural image datasets and evaluating their performance on a new dermatology IQA dataset to discern whether it is possible to use already existing IQA databases without depending on dermatological data. As part of this analysis, we also explored the relationship between input image resolution and model performance: the importance of choosing the right resolution for a deep learning-based IQA model has been previously reported \cite{hosu2020koniq}, and we aimed to extrapolate it to the dermatology domain. Finally, we show the benefits of cross-domain training of IQA models: training simultaneously on many IQA datasets, including dermatology images, provides consistent performance across all natural IQA datasets and improves the performance of dermatology IQA models, as they are trained on a larger pool of image distortions than the ones that could be captured in a smaller, dermatology-only image dataset.

\section{Related work}

\subsection{Image quality control in telemedicine}

There has been an increase in teledermatology guidelines during the last years, many of them including recommendations for dermatological photography \cite{cummins2023consensus}. Despite the availability of this material, image quality is still perceived as the most crucial disadvantage in teledermatology \cite{romero2018modelos}.
Fortunately, thanks to the widespread adoption of digital photography, together with the rapid evolution of hardware, most cameras can fulfill the requirements for good dermatological pictures, which bridges the gap between professionals with and without experience in photography and contributes to a better overall image quality \cite{Pasquali2014PhotographyID}.

However, while healthcare practitioners have shown to be proficient at dermatological photography \cite{dahlen2018teledermoscopy}, patients may struggle at taking pictures. In fact, the review of guidelines conducted in \cite{cummins2023consensus} reveals that dermatological photography recommendations are aimed at healthcare practitioners. This excludes patients from learning how to take good photographs, despite they have become an important player in the clinical process, specially after the COVID-19 pandemic \cite{Tommasino2024}. 

With the widespread adoption of smartphone photography in everyday life, patients are likely to use these devices for teledermatology consultations; although specifications are continuously improving, these devices are not designed with the purpose of dermatology imaging \cite{amouroux2017image}. In addition, patients are most likely to take pictures at their homes, which are non-standardized environments, in non-standardized ways (such as using the front camera of a smartphone), increasing the chances of generating low-quality image data. To overcome these limitations, it is possible to analyze the most critical aspects of medical photography and define some basic principles that clinicians and patients should understand to take good quality pictures, even with a smartphone \cite{zoltie2022medical}. Educating patients for optimal skin photography poses an unsolved challenge, though, as even with proper training they may fail to generate good-quality images \cite{irvine2023ability}. This knowledge gap can even be widened in some populations due to the generational digital divide, as patients of old age are less familiarized with healthcare technology and apps \cite{lopez2022bridging}.

\subsection{IQA methods in dermatology}

The effect of low-quality images on the performance of dermatology AI models has been previously explored by Maier et al. \cite{maier2022image}. Although it is a rather preliminary work limited by their model architecture and distortion choices, it points out that, even when the model is trained with data augmentation to increase robustness, there will always be the risk of encountering either extremely low-quality images or images affected by unexpected distortions, resulting in out-of-domain inputs with inaccurate outputs.

Most research on dermatology IQA has been focused on detecting common artifacts that produce low-quality dermatology pictures. Vodrahalli et al. developed an ensemble of deep learning and classical machine learning models to assess image quality via binary (good/bad) assessments on overall quality, blur, lighting and zoom, but they used a rather small input image size (between 128\textsuperscript{2} and 256\textsuperscript{2} px\textsuperscript{2}) and obtained mixed results \cite{vodrahalli2022development}. Similarly, Jalaboi et al. trained a deep learning model to detect common causes of poor quality (framing, lighting, blur, resolution, and distance) with equally mixed results, even though its performance was comparable to that of their dermatologists and used a much larger dataset \cite{jalaboi2023explainable}. The main disadvantage of both works is that they were designed for explaining what is causing the image to be unsuitable for teledermatology consultations, and did not follow the standardized approach in IQA research, which is rating images using absolute category rating (ACR) scales and measuring correlation between objective scores (the model's predicted quality score) and true subjective scores.

Our previous work \cite{montilla2023dermatology} presented an initial exploration of the usage of non-medical datasets for dermatology IQA, but it lacked a thorough evaluation of every IQA dataset and used a small dermatology image dataset of real distortions (934 images), which may limit models' ability to learn image distortions.

\subsection{IQA databases}

The landscape of IQA datasets outside dermatology IQA is continuously expanding with new image databases. Over the last twenty years, many have been created to characterize the adequateness and performance of quality metrics. The earliest dataset released were IVC \cite{ivcselectencrypt}, LIVE \cite{sheikh2006statistical}, TID2008 \cite{ponomarenko2009tid2008}, and CSIQ \cite{larson2010most}, which consisted of small selections of pristine images to which several distortions were applied. Later on, the TID2008 was later expanded to TID2013 \cite{ponomarenko2015image}, followed by the CID:IQ dataset \cite{liu2014cid}. The CID2013 dataset proposed by Virtanen et al. \cite{virtanen2014cid2013} introduced the novelty of including pictures with concurrent distortions taken with different imaging devices. Ghadiyaram et al. introduced another dataset with authentic distortions, LIVE in the Wild (LIVE-ItW)\cite{ghadiyaram2015massive}, which has become one of the most utilized datasets for benchmarking IQA methods. Their main limitation of these datasets, however, is their small size: as annotating subjective quality scores require large boards of observers, it is a costly and time-consuming process. A more recent dataset, NITS-IQA, provides a larger set of pristine images, but is still limited to a small sample of artificial distortions \cite{ruikar2023nits}.

Recent datasets have addressed this size issue by providing a larger sets of images: Kadid-10k and KADIS-700k \cite{lin2019kadid} provide a large collection of artificially distorted images, while KonIQ-10k \cite{hosu2020koniq}, BIQ2021 \cite{ahmed2022biq2021}, and SPAQ \cite{fang2020perceptual} offer images with naturally occurring distortions.

\section{Materials and methods}

\subsection{Natural IQA datasets}

After reviewing the current landscape of IQA datasets, we selected Kadid-10k \cite{lin2019kadid}, KonIQ-10k \cite{hosu2020koniq}, GFIQA-20k \cite{su2023going}, SPAQ \cite{fang2020perceptual}, BIQ2021 \cite{ahmed2022biq2021}, and LIVE In the Wild \cite{ghadiyaram2015massive} for our experiments. Kadid-10k, KonIQ-10k, GFIQA-20k, SPAQ and BIQ2021 comprise several thousands of images of common everyday scenarios, which made them suitable for our goal of training deep learning models.

The last dataset, LIVE In the Wild, is one of the most common IQA datasets for benchmarking IQA measures and models. Despite its much smaller size, we included it in our study as a separate test set to assess the performance of the models and provide results that are easy to compare to existing literature. A summary of the natural datasets used in this study is presented in Table \ref{natural-datasets-table}.

\begin{table}[]
\caption{Summary of natural IQA datasets used in this work}
\centering
\label{natural-datasets-table}
\begin{tabular}{llll}
\toprule
Dataset   & Number of images & Score range  & Distortion type \\
\midrule
KonIQ-10k & 10073            & {[}1, 5{]}   & Authentic       \\ 
GFIQA-20k & 20000            & {[}0, 1{]}   & Authentic       \\ 
SPAQ      & 11125            & {[}0, 100{]} & Authentic       \\ 
BIQ2021   & 8000             & {[}0, 1{]}   & Authentic       \\ 
LIVE-ItW  & 1169             & {[}0, 100{]} & Authentic       \\ \midrule
Kadid-10k & 10125            & {[}1, 5{]}   & Artificial      \\
\bottomrule
\end{tabular}
\end{table}

\subsection{Dermatology IQA dataset}

\subsubsection{Data collection}

\begin{figure}[ht]
    \centering
    \includegraphics[width=1.00\linewidth]{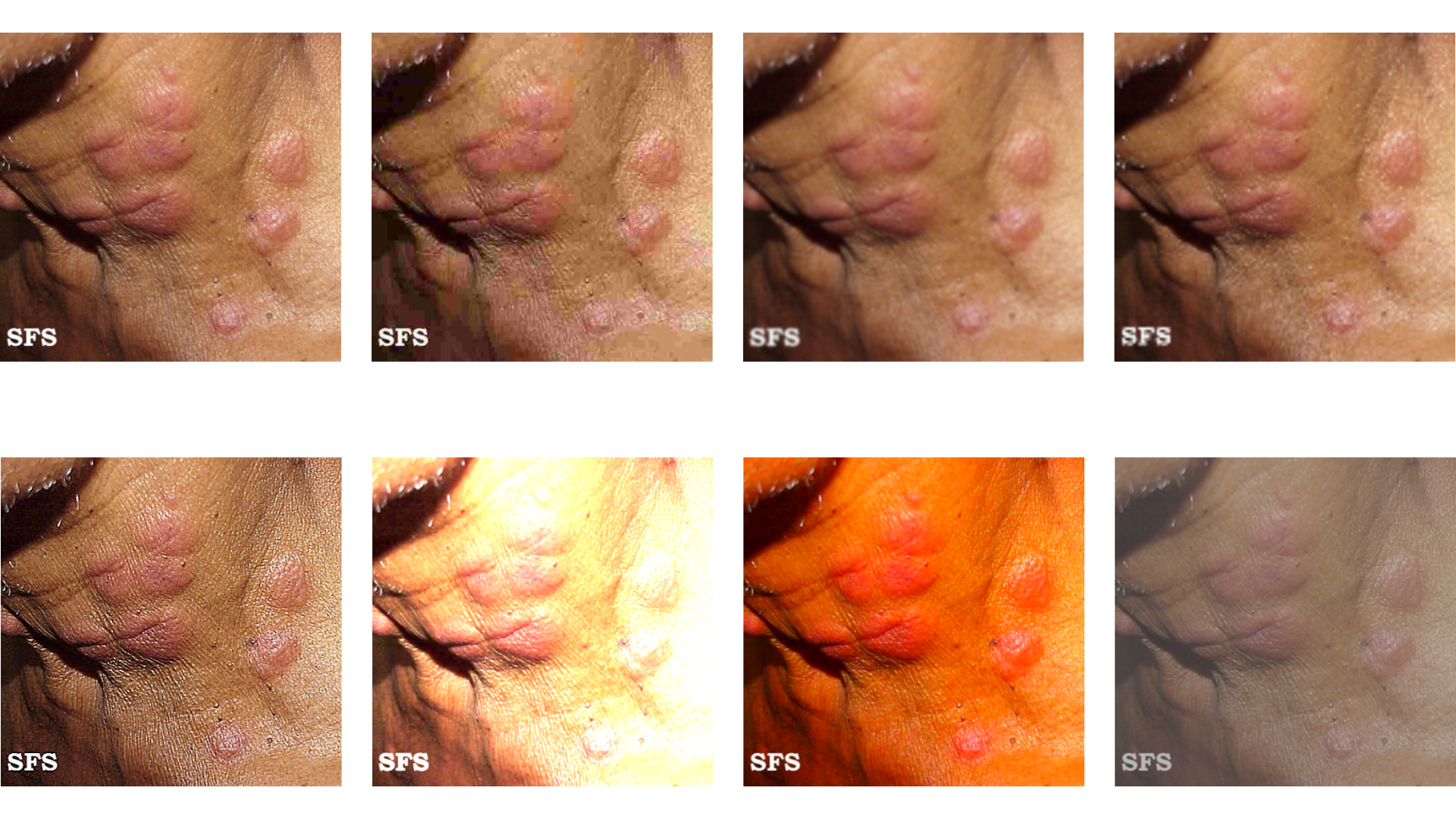}
    \caption{A pristine image (top-left image) of the Legit.Health-DIQA-Artificial dataset and some distorted views. From from top left to bottom right: pristine image, JPEG compression, Gaussian blur, pixelation, sharpening, brightness, color, and contrast distortion. Image source: Atlas Dermatologico (\url{https://www.atlasdermatologico.com.br}).}
    \label{fig:legithealth-diqa-distortions}
    \Description[Some example images of the dermatology IQA dataset created in this study]{An example of a pristine image from the Legit.Health-DIQA-Artificial dataset and some distorted views.}
\end{figure}

For the evaluation of our IQA models on dermatology image quality assessment (DIQA), we created a new dataset, named Legit.Health-DIQA-Artificial. This dataset was generated by selecting 100 pristine dermatology images and applying a set of artificial distortions (Figure \ref{fig:legithealth-diqa-distortions}). Each distortion was applied with different levels of strength, resulting in 1800 additional images in total. The distortions were inspired by that used for Kadid-10k \cite{lin2019kadid}: JPEG compression (3 levels), Gaussian blur (1), pixelation (2), sharpening (3), brightness (4), color (2), and contrast (3).

To create the dataset, we started by collecting images from well-known dermatology atlases and skin image websites (Atlas Dermatologico, Hellenic Dermatological Atlas, Meddean Dermatology Atlas, Black \& Brown Skin). Then, we expanded our search to other skin images datasets, and added PAD-UFES-20 \cite{pacheco2020pad}, ACNE04 \cite{wu2019joint}, PH2 \cite{mendoncca2013ph}, and SMARTSKINS \cite{vasconcelos2014principal}. We also repurposed two biometric datasets related to skin structure: the Transient Biometric Nails datasets \cite{barbosa2013transient}\cite{barros2016use} and the AMI Ear Database \cite{amieardb}. Finally, we expanded the selection to 100 pristine images using with images from internal sources. For most sources, we only used a small portion of images since image content was very similar, whereas dermatology atlases presented higher variability in terms of image content.

A complete list of the image sources can be found in Table \ref{diqa-datasets-table}. By combining images from different sources, we achieved a high variability in terms of acquisition settings, imaging devices, and image content, while ensuring the representation of light and dark skin tones.

\begin{table}[h]
\centering
\caption{Summary of sources used to create the Legit.Health-DIQA-Artificial dataset}
\label{diqa-datasets-table}
\begin{tabular}{lcc}
\toprule
Source                        & Images in pristine selection \\ 
\midrule
Atlas Dermatologico           & 40                           \\ 
Hellenic Dermatological Atlas & 13                           \\ 
Black \& Brown Skin           & 4                            \\ 
Meddean Dermatology Atlas     & 2                            \\ \midrule
PAD-UFES-20                   & 6                            \\ 
ACNE04                        & 4                            \\ 
PH2                           & 3                            \\ 
SMARTSKINS                    & 3                            \\ \midrule
Transient Biometrics Nails    & 1                            \\ 
AMI Ear Database              & 1                            \\ \midrule
Internal                      & 23                           \\ \midrule
\textbf{Total}                & \textbf{100}                          \\
\bottomrule
\end{tabular}
\end{table}

\subsubsection{Data annotation}

The dataset was exclusively annotated for image quality assessment. Each image was rated with a score in the range $[1, 10]$, where 1 and 10 correspond to poor and excellent perceived visual quality, respectively. Prior to annotation, we resized all images so that the largest side was 512 pixels, making them suitable for any display.

\begin{figure}[ht]
    \centering
    \includegraphics[width=1\linewidth]{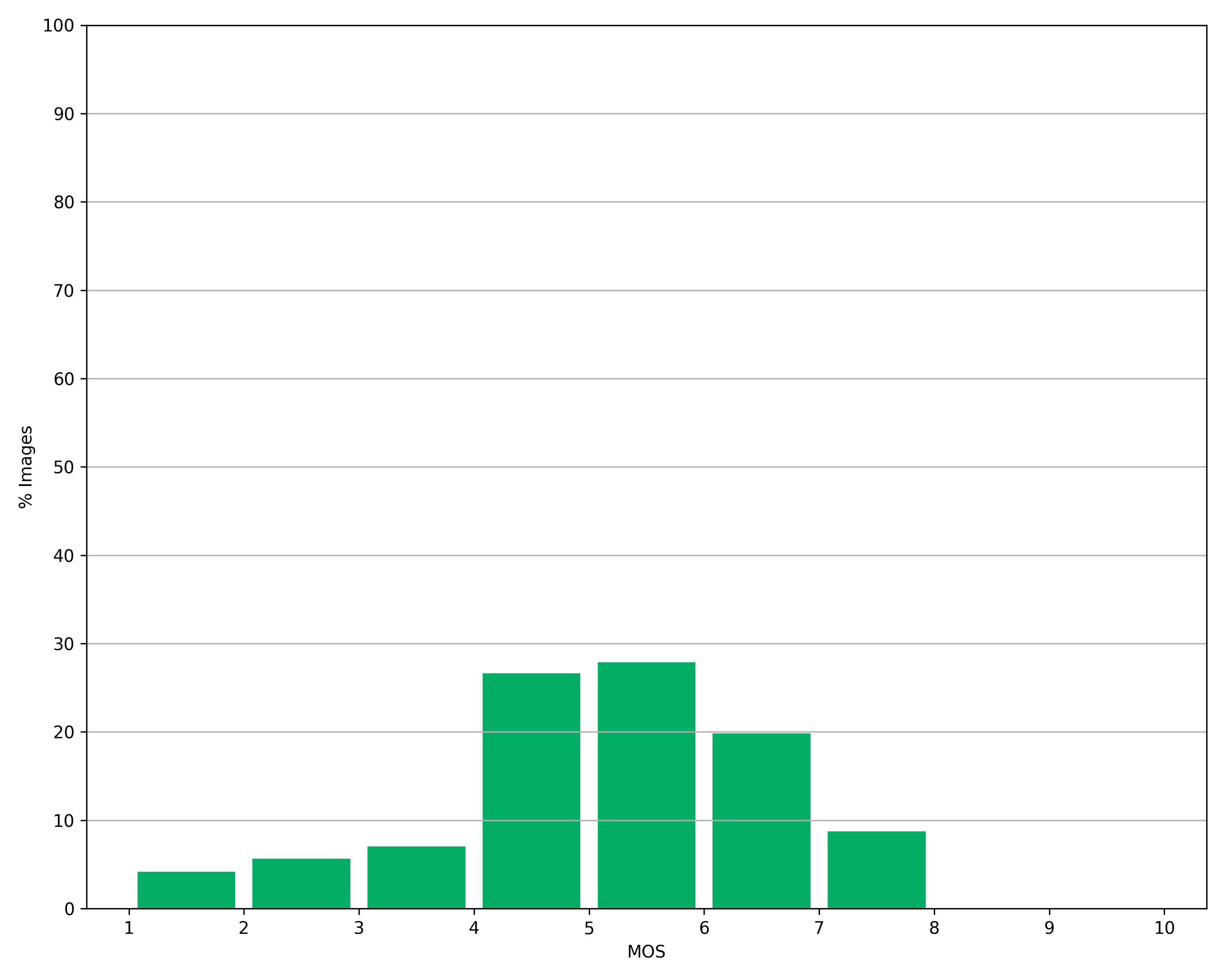}
    \caption{Distribution of the Mean Opinion Scores (MOS) of the Legit.Health-DIQA-Artificial dataset.}
    \label{fig:legithealth-diqa-distributions}
    \Description[Distribution of ratings of the Dermatology IQA dataset]{Histogram of the Mean Opinion Scores (MOS) of the Legit.Health-DIQA-Artificial dataset.}
\end{figure}

We followed the ITU-T P.910 recommendation from the International Telecommunications Union (ITU) and recruited a large board of 40 non-expert observers. This resulted in 40 quality ratings per image, which were averaged to obtain the mean opinion score (MOS). The distribution of MOS for this dataset is presented in Figure \ref{fig:legithealth-diqa-distributions}.

Although this may be considered a medical image dataset, perceptual image quality assessment should not require prior expert knowledge in dermatology, as quality features exist beyond the semantic content of an image. Although perceptual visual quality can be content dependent \cite{siahaan2016does}, we believe that common distortions such as overexposure, underexposure, blur, or compression can be easily detected by any observer in an image, regardless of their medical background.

\subsection{Data preparation}

\begin{figure*}[h]
    \centering
    \includegraphics[width=0.75\linewidth]{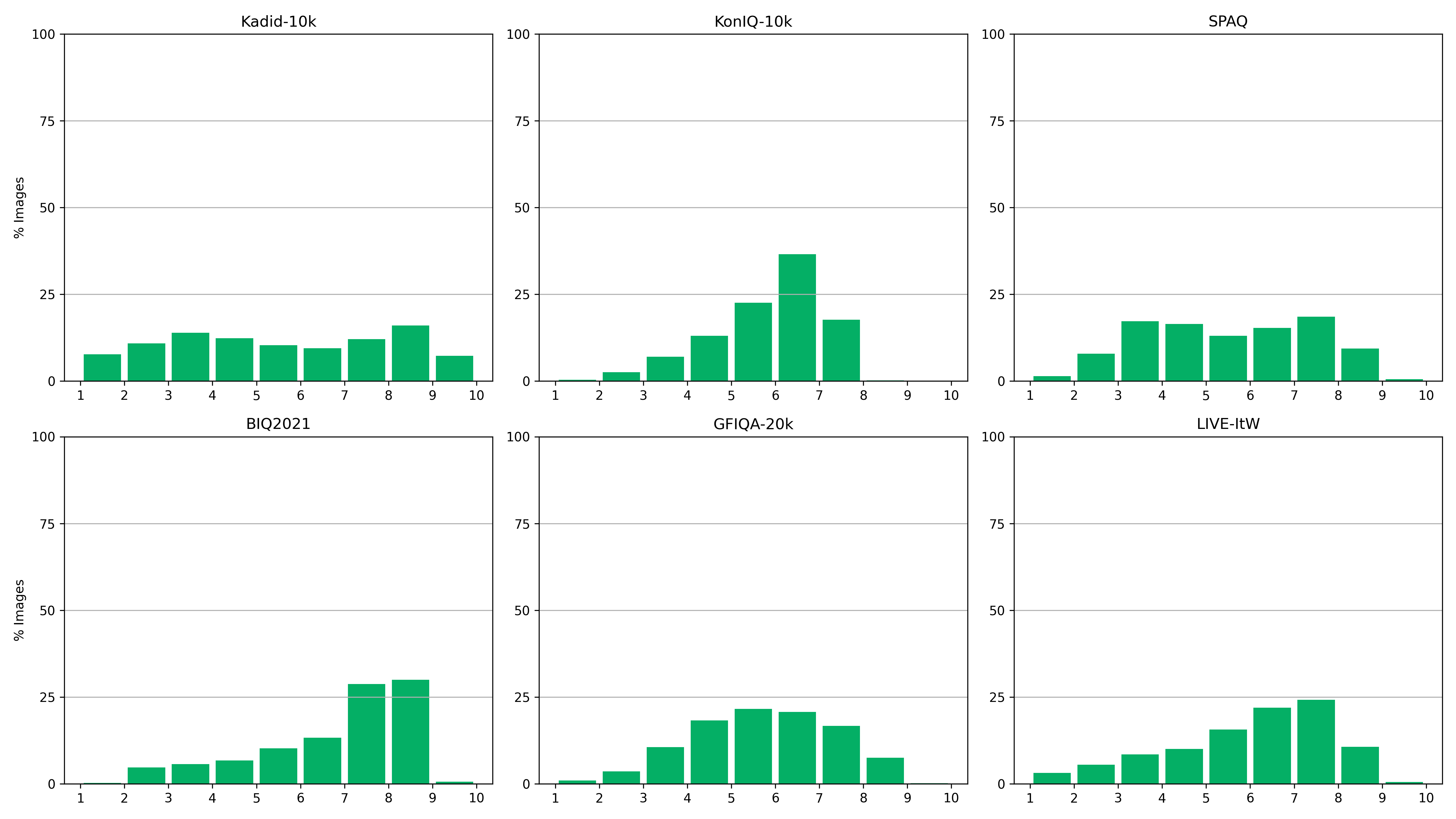}
    \caption{Distribution of Mean Opinion Scores (MOS) for each natural IQA dataset.}
    \label{fig:natural-distributions}
    \Description[Distribution of ratings of the natural IQA datasets]{Histograms of the Mean Opinion Scores (MOS) of the natural IQA datasets.}
\end{figure*}

To make all datasets comparable, we scaled the scores of the natural IQA datasets to the $[1, 10]$ range. The distribution of scores for each natural dataset is presented in figure \ref{fig:natural-distributions}. Lastly, we made the training, validation, and test sets for 5 random splits. For the dermatology images, we were able to group the images manually into subjects during the curation of the Legit.Health-DIQA-Artificial dataset. The entire list of subjects was then divided into five training/validation/test splits; this means that if a pristine image is in the training set, all its distorted views will also be put in that set. This stratification strategy prevented data leakage, which may have led to unreliable results.

For the natural images, we applied different splitting strategies depending on the constraints of each dataset. All image clusters of Kadid-10k (i.e. the pristine image and its distorted views) are independent form each other, which made it possible to divide the dataset into five train/validation/test splits; similarly, all the images from Kadid-10k, KonIQ-10k, GFIQA-20k, and SPAQ are independent samples, so we randomly created five training/validation/test splits. On the contrary, BIQ2021 contained image clusters that were already split into a training and validation set by the original authors, which forced us to use this dataset only for training and validation. Lastly, LIVE In the Wild was always included in the test set of every split, as we stated previously.

\subsection{Model training}

We conducted all the experiments with convolutional neural networks (CNNs), using the EfficientNet architecture. EfficientNet is a family of deep CNNs designed to be both accurate and efficient, using a compound scaling method that uniformly scales the network's width, depth, and input resolution with a fixed ratio. From a base network (EfficientNet-B0), the creators applied such scaling method to generate a collection of models with increasing size and accuracy (up to EfficientNet-B7)\cite{tan2019efficientnet}.

In this work, we used three model sizes (B0, B3, and B5) and adjusted all datasets to their corresponding input sizes ($224^2$, $288^2$, and $448^2$ px\textsuperscript{2}, respectively). We used backbones pre-trained on the ImageNet dataset, adding two fully-connected layers (each with a decreasing number of neurons, a dropout rate of 0.50, and separated by a ReLU activation layer) for the prediction of a single scalar value. Model weights were updated at the training stage using the Mean Squared Error loss function (MSE, eq. \ref{mse_equation}), and validation and test performance was monitored using Pearson's linear correlation coefficient (PLCC, eq. \ref{plcc_equation}). Similarly to other works, we also computed the Spearman's rank correlation coefficient (SROCC, eq. \ref{spearman_equation}) for the test set:

\begin{equation}
\label{mse_equation}
MSE = \frac{1}{n} \sum_{i=1}^{n} (y_i - \hat{y}_i)^2
\end{equation}

\begin{equation}
\label{plcc_equation}
PLCC = \frac{\sum_{i=1}^{n}(x_i - \overline{x})(x_i - \overline{y})}{\sqrt{\sum_{i=1}^{n}(x_i - \overline{x})^2}\sqrt{\sum_{i=1}^{n}(y_i - \overline{y})^2}}
\end{equation}

\begin{equation}
\label{spearman_equation}
SROCC = 1 - \frac{6 \sum_{i=1}^{n} d_i^2}{n(n^2 - 1)}
\end{equation}

\noindent where $x$ and $y$ correspond to the predicted and ground truth scores, respectively, and $d_i$ is the difference between the ranks of predicted and ground truth scores. Both metrics are the most common IQA metrics for measuring the relationship between objective predicted scores and the true subjective scores.

We trained all models for 20 epochs, using the AdamW optimizer \cite{loshchilov2017fixing} with weight decay $\lambda=10^{-5}$ and a one-cycle learning rate policy \cite{smith2019super}. Our learning rate tests, similar to the one proposed in \cite{smith2017cyclical}, showed that a learning rate $\gamma=2\times10^{-4}$ was optimal for all experiments.

To overcome the imbalance of quality ratings, we followed a similar strategy to that used in \cite{bianco2018use}. First, we rounded every quality score, resulting in a set of 10 possible quality levels $l \in \{1\dots10\}$. For a given training dataset $\mathcal{D}$ of images $x_i$ and their corresponding rounded MOS $l_i$, each class weight $w_{l}$ was calculated as:

\begin{equation}
    w_l = \frac{|\mathcal{D}|}{N \cdot |\{x_i \in \mathcal{D} \mid l_i = l\}|}
\end{equation}

\noindent where $N=10$ is the number of levels and $|\mathcal{D}|$ is the size of a training dataset $\mathcal{D}$. Each image was then assigned its corresponding class weight $w_l$ based on its rounded MOS. This method assigned a larger weight to the least represented classes, balancing the training.

Data augmentation is highly restricted in the context of IQA as even the slightest distortion would break the relationship between an image and its MOS label. We followed a similar strategy to the one used to train the NIMA model \cite{talebi2018nima}, which consists of resizing followed by random horizontal flipping and a small amount of random cropping. This does not significantly modify the content of the image but produces some additional variability. In our experiments, we set the initial resizing step to make the image $12.5\%$ larger than the expected input size. At validation and test time, the images are directly resized to the model's input size, followed by a center crop.

For the cross-domain training experiment, we did not require any further processing because the MOS of all datasets were scaled to the same $[1, 10]$ range. With this experiment, we wanted to test if a model is capable of generalizing to all domains given enough examples of good and bad quality images from very different sources.

\section{Results}

The aggregated PLCC and SROCC metrics of all our cross-dataset experiments are presented in Table \ref{results-pearson} and  \ref{results-spearman} (Legit.Health-DIQA-Artificial has been shortened to DIQA for clarity). In terms of average correlation across datasets, the best performing model was the EfficientNet-B5 trained on all datasets. Among all models trained on a single natural IQA dataset, the models trained on the SPAQ dataset achieved the highest performance on the dermatology dataset. As we mentioned in the previous section, the BIQ2021 dataset was only used for training and validation; therefore, we could not get any test metrics on this dataset.

\begin{figure}[h]
    \centering
    \includegraphics[width=1.00\linewidth]{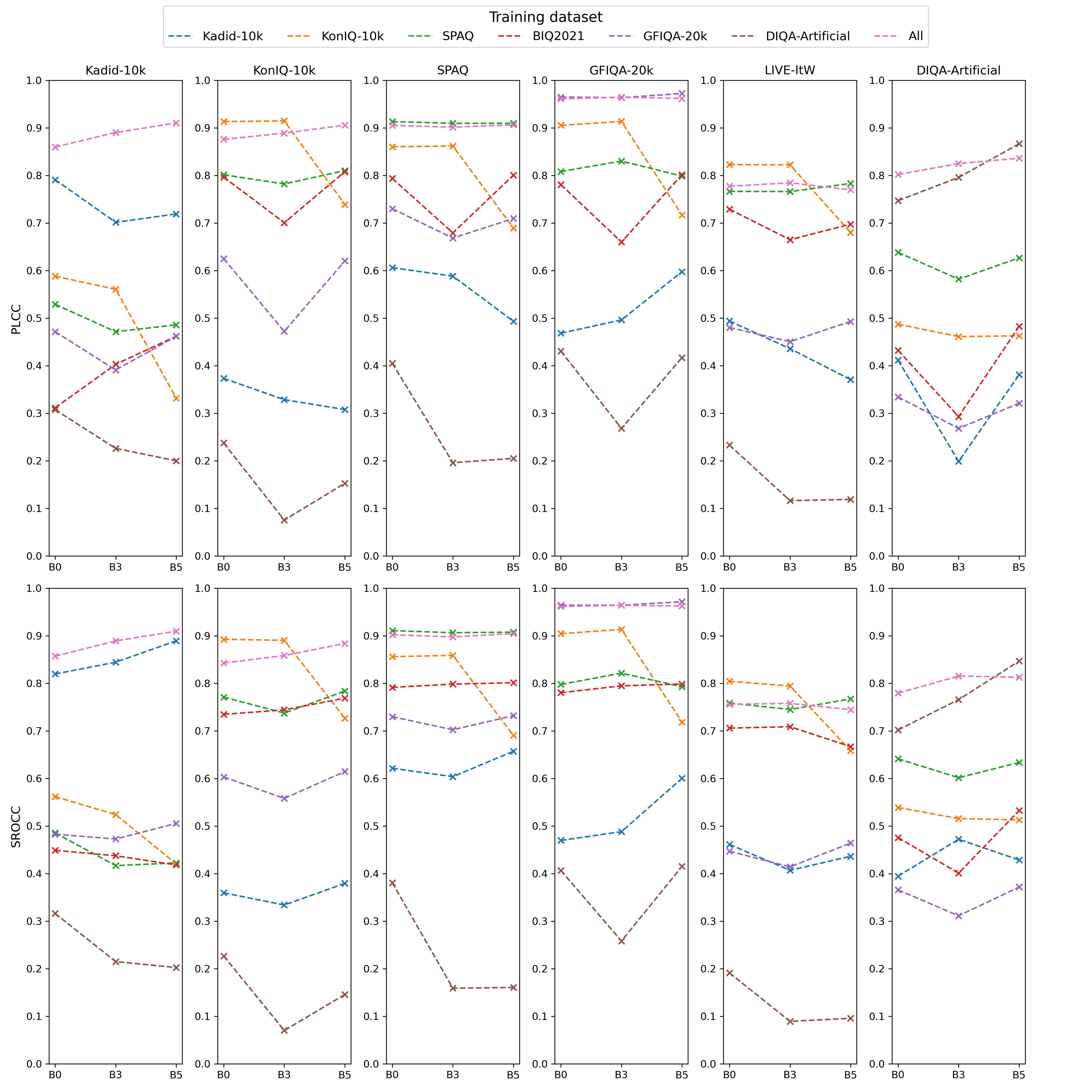}
    \caption{Analysis of the impact of input image size on correlation metrics (PLCC and SROCC).}
    \label{fig:model-size-performance}
    \Description[A plot of performance metrics versus input image size]{A plot of correlation metrics (PLCC and SROCC) versus input image size.}
\end{figure}

The effect of using bigger input sizes is presented in Figure \ref{fig:model-size-performance}. Based on these results, we observed that the effect is heavily dataset-dependent, with no clear overall benefit shown when using larger models and single-domain training. However, for cross-domain training, using larger models led to better performance on most datasets.

\begin{table*}[]
\caption{Pearson linear correlation coefficients of the IQA models on all test sets (mean ± standard deviation).}
\label{results-pearson}
\centering
\begin{tabular}{llllllll}
\toprule
Model & Dataset                     & Kadid-10k                & KonIQ-10k                & SPAQ                     & GFIQA-20k                & LIVE-ItW                 & DIQA \\
\midrule
B0    & Kadid-10k                   & 0.7911 ± 0.0200          & 0.3733 ± 0.0567          & 0.6060 ± 0.0400          & 0.4684 ± 0.0548          & 0.4939 ± 0.0162          & 0.4116 ± 0.0951             \\
B3    & Kadid-10k                   & 0.7014 ± 0.1682          & 0.3286 ± 0.0557          & 0.5883 ± 0.0160          & 0.4962 ± 0.0469          & 0.4359 ± 0.0118          & 0.1989 ± 0.2543             \\
B5    & Kadid-10k                   & 0.7191 ± 0.3770          & 0.3078 ± 0.2213          & 0.4933 ± 0.3058          & 0.5977 ± 0.0527          & 0.3708 ± 0.2427          & 0.3813 ± 0.0602             \\ \hline
B0    & KonIQ-10k                   & 0.5881 ± 0.0170          & \bftab 0.9134 ± 0.0061   & 0.8602 ± 0.0038          & 0.9054 ± 0.0046          & \bftab 0.8228 ± 0.0067 & 0.4872 ± 0.0518             \\
B3    & KonIQ-10k                   & 0.5608 ± 0.0200          & \bftab 0.9146 ± 0.0033 & 0.8620 ± 0.0048          & 0.9136 ± 0.0034          & \bftab 0.8223 ± 0.0065 & 0.4613 ± 0.0415             \\
B5    & KonIQ-10k                   & 0.3316 ± 0.3343          & 0.7390 ± 0.4266          & 0.6894 ± 0.4016          & 0.7167 ± 0.4475          & 0.6802 ± 0.3679          & 0.4629 ± 0.2244             \\ \hline
B0    & SPAQ                        & 0.5291 ± 0.0428          & 0.8014 ± 0.0072          & \bftab 0.9133 ± 0.0022 & 0.8084 ± 0.0208          & 0.7666 ± 0.0037          & 0.6383 ± 0.0461             \\
B3    & SPAQ                        & 0.4715 ± 0.0416          & 0.7821 ± 0.0093          & \bftab 0.9095 ± 0.0015 & 0.8302 ± 0.0133          & 0.7661 ± 0.0051          & 0.5819 ± 0.0336             \\
B5    & SPAQ                        & 0.4859 ± 0.0463          & 0.8103 ± 0.0052          & \bftab 0.9097 ± 0.0009 & 0.7987 ± 0.0144          & 0.7832 ± 0.0301          & 0.6267 ± 0.0786             \\ \hline
B0    & GFIQA-20k                   & 0.4715 ± 0.0374          & 0.6251 ± 0.0420          & 0.7299 ± 0.0278          & \bftab 0.9648 ± 0.0010 & 0.4803 ± 0.0225          & 0.3342 ± 0.0900             \\
B3    & GFIQA-20k                   & 0.3911 ± 0.1333          & 0.4722 ± 0.1929          & 0.6684 ± 0.0370          & 0.9636 ± 0.0031          & 0.4507 ± 0.0349          & 0.2682 ± 0.0983             \\
B5    & GFIQA-20k                   & 0.4617 ± 0.0432          & 0.6204 ± 0.0211          & 0.7096 ± 0.0383          & \bftab 0.9723 ± 0.0005 & 0.4930 ± 0.0588          & 0.3211 ± 0.0536             \\ \hline
B0    & DIQA & 0.3074 ± 0.0632          & 0.2376 ± 0.0459          & 0.4052 ± 0.0437          & 0.4307 ± 0.0684          & 0.2330 ± 0.0322          & 0.7470 ± 0.0525             \\
B3    & DIQA & 0.2261 ± 0.0287          & 0.0754 ± 0.0452          & 0.1961 ± 0.0778          & 0.2682 ± 0.0489          & 0.1164 ± 0.0581          & 0.7958 ± 0.0393             \\
B5    & DIQA & 0.2002 ± 0.0747          & 0.1528 ± 0.0700          & 0.2051 ± 0.1348          & 0.4170 ± 0.0983          & 0.1189 ± 0.0745  & \bftab 0.8666 ± 0.0400             \\ \hline
B0    & All                         & \bftab 0.8597 ± 0.0106 & 0.8759 ± 0.0086          & 0.9052 ± 0.0025          & 0.9610 ± 0.0017          & 0.7773 ± 0.0113          & 0.8021 ± 0.0550             \\
B3    & All                         & \bftab 0.8903 ± 0.0096 & 0.8890 ± 0.0040          & 0.9015 ± 0.0029          & \bftab 0.9646 ± 0.0009 & \bftab 0.7845 ± 0.0044 & \bftab 0.8249 ± 0.0237             \\
B5    & All                         & \bftab 0.9102 ± 0.0086 & \bftab 0.9056 ± 0.0069 & 0.9064 ± 0.0027          & 0.9619 ± 0.0023          & 0.7700 ± 0.0141          & \bftab 0.8363 ± 0.0349    \\
\bottomrule
\end{tabular}
\end{table*}

\begin{table*}[]
\caption{Spearman's rank correlation coefficients of the IQA models on all test sets (mean ± standard deviation). Top-3 best models for each dataset are bolded.}
\label{results-spearman}
\centering
\begin{tabular}{llllllll}
\toprule
Model & Dataset                     & Kadid-10k                & KonIQ-10k                & SPAQ                     & GFIQA-20k                & LIVE-ItW                 & DIQA \\
\midrule
B0    & Kadid-10k                   & 0.8195 ± 0.0202          & 0.3594 ± 0.0581          & 0.6213 ± 0.0422          & 0.4696 ± 0.0564          & 0.4611 ± 0.0175          & 0.3945 ± 0.1072             \\
B3    & Kadid-10k                   & 0.8444 ± 0.0216          & 0.3338 ± 0.0244          & 0.6037 ± 0.0140          & 0.4881 ± 0.0457          & 0.4068 ± 0.0117          & 0.4721 ± 0.0558             \\
B5    & Kadid-10k                   & \bftab 0.8893 ± 0.0156 & 0.3794 ± 0.0623          & 0.6571 ± 0.0292          & 0.6003 ± 0.0432          & 0.4363 ± 0.0285          & 0.4286 ± 0.0726             \\ \hline
B0    & KonIQ-10k                   & 0.5618 ± 0.0222          & \bftab0.8925 ± 0.0067 & 0.8560 ± 0.0048          & 0.9043 ± 0.0052          & \bftab 0.8043 ± 0.0066 & 0.5389 ± 0.0455             \\
B3    & KonIQ-10k                   & 0.5240 ± 0.0230          & \bftab 0.8903 ± 0.0063 & 0.8587 ± 0.0046          & 0.9133 ± 0.0037          & \bftab 0.7942 ± 0.0070 & 0.5156 ± 0.0352             \\
B5    & KonIQ-10k                   & 0.4203 ± 0.2628          & 0.7264 ± 0.4205          & 0.6905 ± 0.4022          & 0.7184 ± 0.4445          & 0.6583 ± 0.3562          & 0.5128 ± 0.2363             \\ \hline
B0    & SPAQ                        & 0.4855 ± 0.0400          & 0.7710 ± 0.0097          & \bftab 0.9107 ± 0.0034 & 0.7976 ± 0.0249          & 0.7586 ± 0.0053          & 0.6412 ± 0.0453             \\
B3    & SPAQ                        & 0.4168 ± 0.0412          & 0.7374 ± 0.0231          & \bftab 0.9064 ± 0.0029 & 0.8213 ± 0.0158          & 0.7450 ± 0.0058          & 0.6015 ± 0.0375             \\
B5    & SPAQ                        & 0.4223 ± 0.0557          & 0.7837 ± 0.0102          & \bftab 0.9075 ± 0.0027 & 0.7928 ± 0.0170          & \bftab 0.7672 ± 0.0327 & 0.6335 ± 0.0702             \\ \hline
B0    & GFIQA-20k                   & 0.4827 ± 0.0276          & 0.6031 ± 0.0391          & 0.7296 ± 0.0292          & \bftab 0.9643 ± 0.0011 & 0.4469 ± 0.0206          & 0.3657 ± 0.0984             \\
B3    & GFIQA-20k                   & 0.4727 ± 0.0213          & 0.5579 ± 0.0483          & 0.7024 ± 0.0290          & \bftab 0.9643 ± 0.0010 & 0.4140 ± 0.0342          & 0.3112 ± 0.1010             \\
B5    & GFIQA-20k                   & 0.5054 ± 0.0334          & 0.6145 ± 0.0224          & 0.7318 ± 0.0327          & \bftab 0.9715 ± 0.0006 & 0.4640 ± 0.0657          & 0.3716 ± 0.0672             \\ \hline
B0    & DIQA & 0.3162 ± 0.0458          & 0.2266 ± 0.0511          & 0.3807 ± 0.0486          & 0.4065 ± 0.0733          & 0.1912 ± 0.0443          & 0.7019 ± 0.0500             \\
B3    & DIQA & 0.2149 ± 0.0270          & 0.0701 ± 0.0494          & 0.1589 ± 0.0789          & 0.2582 ± 0.0557          & 0.0892 ± 0.0543          & 0.7657 ± 0.0383             \\
B5    & DIQA & 0.2024 ± 0.0710          & 0.1456 ± 0.0839          & 0.1605 ± 0.1430          & 0.4151 ± 0.1015          & 0.0958 ± 0.0760          & \bftab 0.8466 ± 0.0420    \\ \hline
B0    & All                         & 0.8570 ± 0.0108          & 0.8427 ± 0.0064          & 0.9020 ± 0.0037          & 0.9614 ± 0.0017          & 0.7559 ± 0.0152          & 0.7793 ± 0.0600             \\
B3    & All                         & \bftab 0.8891 ± 0.0115 & 0.8584 ± 0.0072          & 0.8975 ± 0.0043          & 0.9639 ± 0.0008          & 0.7578 ± 0.0066          & \bftab 0.8153 ± 0.0225    \\
B5    & All                         & \bftab 0.9096 ± 0.0107 & \bftab 0.8838 ± 0.0109 & 0.9045 ± 0.0049          & 0.9625 ± 0.0019          & 0.7445 ± 0.0138          & \bftab 0.8125 ± 0.0364    \\
\bottomrule
\end{tabular}
\end{table*}

\section{Discussion}

Our results are consistent with previous cross-dataset evaluations of IQA models, which showed that IQA models trained on one dataset did not perform well when tested on others \cite{yang2019cnn}\cite{lin2020deepfl}. We have discovered that this also occurs when cross-testing on medical image datasets. However, our work presents a limitation similar to previous attempts at cross-domain training, which is not being able to perform MOS scaling. Without proper and dataset-specific scaling, images from different datasets with significantly different quality features may be scaled to a similar MOS, which can reduce the effectiveness of combining IQA datasets \cite{zhang2021uncertainty}. In future developments, we will include images from several IQA datasets to enable MOS scaling for a more precise training and validation, instead of naively rescaling all datasets to $[1, 10]$. 

In our single-domain experiments, the models trained on the SPAQ dataset yielded the best individual performance on the Legit.Health-DIQA-Artificial dataset, showing strong correlation (PLCC and SROCC greater than 0.6).

The models trained on the Kadid-10k dataset ranked the lowest positions in terms of correlation on the dermatology dataset; conversely, the models trained on Legit.Health-DIQA-Artificial performed poorly on Kadid-10k. We find these results surprising since they are the only two artificial datasets used in this study, and they contain similar distortions. This may suggest that the models learned the image content instead of the artificial distortions, leading to poor generalization. This may also be happening to the models trained on KonIQ-10k, SPAQ, and GFIQA-20K, as they score the highest correlation metrics in their corresponding test sets.

Another cause of the domain gap may have been the quality rating procedure of each dataset, as perceived visual quality is content dependent to some extent \cite{siahaan2016does}. In dermatology IQA, we suspect that image quality is deeply entangled with visual aesthetics and clinical utility. When rating dermatology images, annotators may be influenced by the underlying purpose of the images, which is to capture a disease or a lesion; inner judgment of how well the image captures such clinical information may be affecting the rating of visual quality features. This behavior would explain why not a single image from the dataset obtained a MOS higher than 8 (Figure \ref{fig:legithealth-diqa-distributions}). Moreover, visual aesthetics of medical images are entirely different form a natural image, which may be adding some difficulty to this task for non-expert observers, even after receiving proper training.

Despite these limitations, we conclude that cross-domain training of an IQA model for dermatology IQA tasks provides better results than training exclusively on dermatology data. In tables \ref{results-pearson} and \ref{results-spearman}, we can observe that the performance of the cross-domain models on the DIQA dataset is superior in most cases, with the exception of the EfficientNet-B5 variants, which may be suffering from overfitting.

In conclusion, we believe that it is not reliable to use non-dermatology IQA models individually for the assessment of teledermatology images. Instead, models should be trained using a cross-domain approach, including curated dermatology IQA datasets, preferably annotated by dermatologists.

\section{Conclusion}

In this work, we covered the use of clinical and non-clinical image data to overcome the main limitations of current approaches in dermatology image quality assessment (DIQA), namely limited training data and nonstandard definitions of perceived visual quality assessment. We followed the common approach for IQA development, leveraging not only an already validated methodology but also a variety of natural IQA databases released in recent years.

We combined the existing natural IQA datasets with a novel artificially distorted dermatology image quality database, Legit.Health-DIQA-Artificial, annotated by non-expert human observers. In future works, we will explore the differences between the image quality ratings of non-expert human observers and dermatologists to discern whether it is mandatory to obtain quality ratings from expert observers or human crowds suffice, as well as more precise methods of MOS scaling between datasets. We will also leverage dermatology IQA datasets with authentic distortions to account for real-life scenarios.

In conclusion, cross-domain training of IQA models expands their capabilities and ensure proper performance across image domains, distortions, and observer groups. This innovation allows for seamless implementation of DIQA in teledermatology workflows, ensuring visual quality and maximizing the clinical utility of images. As a result, DIQA modles not only have the potential to improve the reliability of daily clinical practice but also to strengthen the quality and consistency of dermatological data used in clinical trials, ultimately benefiting patients and advancing dermatological research.

\section{Data availability statement}

The Legit.Health-DIQA-Artificial dataset image quality ratings and code are not publicly available due to intellectual property restrictions.

Online dermatology images were collected from Atlas Dermatologico \url{https://atlasdermatologico.com.br/}, Meddean Dermatology Atlas \url{https://www.meddean.luc.edu/lumen/meded/medicine/dermatology/melton/atlas.htm}, Black \& Brown skin \url{https://www.blackandbrownskin.co.uk/campaigns}, and Hellenic Dermatological Atlas \url{https://www.hellenicdermatlas.com/en/search/browse}.

\begin{acks}
This project has been funded by the Department of Industry, Energy Transition and Sustainability of the Basque Government (BIKAINTEK Program). We also thank LinkedAI for their data annotation services.
\end{acks}

\bibliographystyle{ACM-Reference-Format}
\bibliography{cz0002-icbra2025}

\appendix









\end{document}